\documentclass[aps,prc,twocolumn,superscriptaddress,showpacs,floatfix]{revtex4-1}
\usepackage{graphicx,amsmath,amssymb,bm}

\usepackage{braket}
\usepackage{amsfonts}

\newcommand{\anntilde}[1]{\tilde a_{#1}}
\newcommand*{\intsum}{\mathop{\ooalign{\raise0.2pt\hbox{$\int$}%
            \cr\lower0.3pt\hbox{$\Sigma$}}}}

\begin{document}

\title{Closed-shell properties of $^{24}$O with {\em ab initio} coupled-cluster theory}

\author{\O.~Jensen}
\affiliation{Department of Physics and Technology, University of Bergen, N-5007 Bergen, Norway}
\affiliation{Department of Physics and Center of Mathematics for Applications, University of Oslo, N-0316 Oslo, Norway}
\author{G.~Hagen}
\affiliation{Physics Division, Oak Ridge National Laboratory,
Oak Ridge, TN 37831, USA}
\affiliation{Department of Physics and Astronomy, University of
Tennessee, Knoxville, TN 37996, USA}
\author{M.~Hjorth-Jensen}
\affiliation{Department of Physics and Center of Mathematics for Applications, University of Oslo, N-0316 Oslo, Norway}
\author{J.~S.~Vaagen}
\affiliation{Department of Physics and Technology, University of Bergen,
N-5007 Bergen, Norway}

\begin{abstract}

  We present an \emph{ab initio} calculation of spectroscopic factors for
  neutron and proton removal from $^{24}$O using the coupled-cluster
  method and a state-of-the-art chiral nucleon-nucleon interaction at
  next-to-next-to-next-to-leading order.
  In order to account for the coupling to the scattering continuum we use a
  Berggren single-particle basis that treats bound, resonant, and continuum states on
  an equal footing. We report neutron removal spectroscopic factors for
  the $^{23}$O states $J^{\pi} = 1/2^+$, $5/2^+$, $3/2^-$ and
  $1/2^-$, and proton removal spectroscopic factors for the $^{23}$N states $1/2^-$ and $3/2^-$.
  Our calculations support the accumulated experimental evidence that $^{24}$O is a closed-shell nucleus.

\end{abstract}


\maketitle

\emph{Introduction \label{secIntro}}
The study of nuclei far from stability is a  leading direction in
nuclear physics, experimentally and theoretically. It represents a
considerable intellectual challenge to our understanding of the
stability of matter itself, with potential implications for the
synthesis of elements. An important aspect of this research direction is
to understand how magic numbers and shells appear and evolve with
increasing numbers of neutrons or protons. The structure and properties
of barely stable nuclei at the limits of stability, have been
demonstrated to deviate dramatically from the established
patterns for ordinary and stable nuclei, see
Ref.~\cite{Thoennessen2004} and references therein for a recent review.
One of the striking features of
nuclei close to the drip line is the adjustement of shell gaps, giving rise to
different {\em magic numbers} \cite{Ozawa2000}. The way shell closures
and single-particle properties evolve as functions of the
number of nucleons forms one of the greatest challenges
to our understanding of the basic features of nuclei, and thereby the
stability of matter.

The chain of oxygen isotopes up to $^{28}$O is particularly
interesting since these are the heaviest nuclei for which the drip
line is well established. Two out of four stable even-even
isotopes exhibit a doubly magic nature, namely $^{22}$O ($Z$=$8$,$N$=$14$)
and $^{24}$O ($Z$=$8$, $N$=$16$). Several recent experiments
\cite{Kanungo2009,stanoiu2004,hoffman2009,catford2010} bring evidence that $^{24}$O is
the last stable oxygen isotope. This is remarkable, in
particular if one considers the fact that the addition of a
single proton on top of the $Z=8$ closed shell brings the drip line
of the fluorine isotopes to $^{31}$F. Recent measurements \cite{Kanungo2009}
also suggest that $^{24}$O has a spherical neutron configuration. The isotopes $^{25-28}$O
are all believed to be unstable towards neutron emission, even though $^{28}$O
is a doubly magic nucleus within the standard shell-model picture. This indicates
that the magic number at the neutron drip line for the oxygen isotopes
is not at  $N=20$ but rather at $N=16$.

Although spectroscopic factors are not observable quantities
\cite{Furnstahl2002, Furnstahl2010}, they can be used to address shell
closure properties within the context of a given model. Experimentally,
spectroscopic factors are defined as the ratio of the observed reaction rate
with respect to the same rate, calculated with a particular reaction model and
assuming a full occupation of the relevant single-particle states. The
spectroscopic factors are then interpreted as the occupancy of specific
single-particle states. Theoretically however, spectroscopic factors measure
what fraction of the full wave function corresponds to the product of a
correlated state (often chosen to be a given closed-shell core) and an
independent single-particle or single-hole state. Large deviations from the
values predicted by an independent-particle model point to a strongly
correlated system.

Kanungo {\em et al} \cite{Kanungo2009} reported measurements
of one-neutron removal from $^{24}$O, and an extraction of spectroscopic
factors. They used an eikonal reaction model and Woods-Saxon overlap functions,
that were calculated with a well depth adjusted to reproduce the one-neutron separation
energies. The results were compared with shell model predictions
using the fitted $sd$-shell USDB interaction \cite{brown2006} and the
SDPF-M interaction \cite{utsuno1999}. The theoretical calculations
corroborate the large $s$-wave probability found in the experimental
analysis, implying thereby that $^{24}$O is indeed a doubly magic nucleus.

The aim of this work is to add further theoretical evidence and
support to these claims. We present spectroscopic
factors for neutron and proton removal from $^{24}$O as predicted by
the \emph{ab initio} coupled-cluster method with the chiral
nucleon-nucleon interaction by Entem and Machleidt~\cite{Entem2003} at
next-to-next-to-next-to-leading order (N$^3$LO). Our calculations are
performed in a large single-particle basis which includes
bound and continuum single-particle states. The above mentioned
theoretical calculations of Refs.~\cite{brown2006,utsuno1999} involve
only fitted effective interactions tailored to small shell-model spaces,
such as the $sd$ or the $sd-pf$ shells only.

The virtue of \emph{ab initio} methods applied to nuclear physics is
to reduce the model dependence of computed results. By means of a
recipe for systematic improvements, one can distinguish between
parameters of technical and physical character. Whereas the result may be
expected to depend on physical parameters, it should be insensitive to the
technical parameters as systematic improvements are included. Ultimately, a
converged \emph{ab-initio} result may provide a rigorous test of the nuclear
interaction model and the corresponding physical parameters.
Since our single-particle basis contains continuum states, and therefore
represents the correct asymptotical behaviour, we are able to generate
radial overlap functions for drip-line nuclei. This work paves therefore the way
for neutron-knockout reactions with fully \emph{ab initio} structure
information.

After these introductory remarks, we briefly expose our calculational
formalism as well as the choice of Hamiltonian and basis in the next
sections.  Following that, our results are presented.
Conclusions and perspectives are drawn in the final section.

\emph{Method}
The aim of this section is to give a short overview of the
steps in our calculations.  The coupled-cluster method
\cite{Cizek1969,Cizek1966, Bartlett2007,
Coester1958, Coester1960, dean2004, HPD+08,Hagen2009a, Hagen2010,
  Hagen2010a, Hagen2007} and the application to spectroscopic
factors \cite{Jensen2010} have been presented in great detail
elsewhere. Here we give only a brief summary of the concepts that
enter the calculations leading to the results presented in this article.

The spectroscopic factor (SF) is the norm of the overlap function,
\begin{align}
	S_{A-1}^A(lj) &= \left|O_{A-1}^A(lj;r)\right|^2 \label{equDefSF}\,,\\
	O_{A-1}^A(lj;r) &= \intsum_n\Braket{A-1||\anntilde{nlj}||A}\phi_{nlj}(r)\,.
	\label{equDefOverlap}
\end{align}
Here, $O_{A-1}^A(lj;r)$ is the radial overlap function of the many-body
wavefunctions for the two independent systems with $A$ and $A-1$
particles respectively.  The double bar denotes a reduced matrix
element, and the integral-sum over $n$ represents both the sum over the
discrete spectrum and an integral over the corresponding continuum part of the spectrum. The annihilation
operator $\anntilde{nljm}$ transforms like a spherical tensor of rank
$j$ and projection $m$.  The radial single particle
basis function is given by the term $\phi_{nlj}(r)$, where $l$ and $j$ denote the single particle orbital and
angular momentum, respectively, and $n$ is the nodal quantum number. The isospin
quantum number has been suppressed.  We emphasize
that the overlap function, and hence also its norm, is defined
microscopically and independently of the single particle basis.  It is
uniquely determined by the many-body wave functions $\ket{A}$ and
$\ket{A-1}$. The quality of the SF estimate is thus limited by the
quality of the pertinent many-body wave functions.

Our calculation of spectroscopic factors follows the recipe detailed in
Ref.~\cite{Jensen2010}.  The important difference between this work and
Ref.~\cite{Jensen2010}, is that
all terms contributing to the spectroscopic factors have now been
expressed in terms of reduced matrix elements in an angular momentum coupled
basis. This allows us to handle  a much larger set of single-particle states as discussed
in Refs.~\cite{HPD+08,Hagen2010a}.

We use the coupled-cluster ansatz, $\ket{\psi_0} = \exp{(T)}\ket{\phi_0}$
for the ground state of the closed-shell nucleus $^{24}$O.
The reference state, $\ket{\phi_0}$, is an antisymmetric product state
for all $A$ nucleons.  The cluster operator $T$ introduces correlations
as a linear combination of particle-hole excitations $T = T_1 + T_2 +
\ldots + T_A$, where $T_n$ represents an $n$-particle-$n$-hole
excitation operator.  For the coupled-cluster singles and doubles
approximation (CCSD) employed in this work, $T$ is truncated at the
level of double excitations, $T = T_1 + T_2$. The coupled-cluster
solution for $^{24}$O is obtained as a set of amplitudes that defines
$T$.

Due to the non-hermiticity of the standard coupled-cluster ansatz, we need both the left eigenvectors and
the right eigenvectors. These are determined via
the equation-of-motion coupled-cluster (EOM-CC) approach as
$\ket{A} \approx \ket{R_\nu^{A}(J_{A})}\equiv \exp{(T)} R_\nu^{A}(J_{A}) \ket{\phi_0}$
and $\bra{A} \approx \bra{L_\nu^{A}(J_{A})} \equiv \bra{\phi_0} L_{\nu}^{A}(J_A) \exp{(-T)}$.
The operators $R_\nu^A(J_A)$ and $L_\nu^A(J_A)$ produce linear
combinations of particle-hole excited states when acting to
the right and left, respectively.
In the spherical form of the EOM-CC approach, the
operators have well defined angular
momentum by construction, as indicated by $J_A$, which stands for the angular
momentum considered.  See for example Ref.~\cite{Hagen2010a}.
If the $A$-body system is in its
ground state, the right EOM-CC wave function is identical to the
coupled-cluster ground state.

Solutions for the $A-1$-body system, $^{23}$O and $^{23}$N, are
obtained with particle-removal equations-of-motion (PR-EOM-CCSD),
where we use the CCSD ground state solution of $^{24}$O as the
reference state in order to determine the corresponding left and right eigenvectors as $\ket{A-1} \approx \ket{R_\mu^{A-1}(J_{A-1})} \equiv \exp{(T)} R_\mu^{A-1}(J_{A-1}) \ket{\phi_0}$ and $\bra{A-1} \approx\bra{L_\mu^{A-1}(J_{A-1})} \equiv\bra{\phi_0} L_\mu^{A-1}(J_{A-1}) \exp{(-T)}$.
In actual calculations, the EOM-CC wave functions are obtained by determining
the operators $R_\nu^A(J_A)$ and $L_\nu^A(J_A)$ as eigenvectors of the
similarity transformed Hamiltonian, $\overline H=\exp{(-T)}H\exp{(T)}$.  The
transformed Hamiltonian is non-hermitian, implying that the left eigenstates must be
determined independently.  We refer the reader to
Refs.~\cite{Bartlett2007, Hagen2010} for details about the
equation-of-motion approach combined with coupled-cluster theory.

Finally, we can approximate the spectroscopic factor,
Eq.~(\ref{equDefSF}) in the spherical coupled-cluster formalism as
\begin{align}
S_{A-1}^{A}(lj) &= \intsum_n
	\braket{L^{A-1}_\mu(J_{A-1})|| \overline{\tilde a_{nlj}} || R^A_\nu(J_A)}
	\nonumber\\\times
	&\braket{R^{A-1}_\mu(J_{A-1})|| \overline{\tilde a_{nlj}} || L^A_\nu(J_A)}^*,
	\label{equSFdefinition}
\end{align}
where we have used the similarity transformed spherical annihilation operator as
\[\overline{\tilde a_{nljm}} = \exp{(-T)}\tilde a_{nljm}\exp{(T)}.\]
Closed expressions for the similarity transformed operators  are given
in Ref.~\cite{Jensen2010}. The labels $\mu$ and
$\nu$ are included to distinguish excited states of the (PR-)EOM-CC solutions. In the spherical
formulation of EOM-CCSD, the solutions are spherical tensors
\cite{Hagen2010a}, and the spectroscopic factor depends on the rank, but not on
the projection of the EOM-CCSD states. In order to derive the coupled
expressions, a Racah algebra module was developed for the open source computer
algebra system SymPy \cite{sympy}.  More details about these calculations can be
found in Ref.~\cite{Phdjensen2010}.

\emph{Hamiltonian and Basis}
We use an intrinsic $A$-nucleon Hamiltonian $\hat{H} = \hat{T}-\hat{T}_{\rm cm} +\hat{V}$
where $\hat T$ is the kinetic energy, $\hat T_{\rm cm}$ is the
kinetic energy of the center-of-mass coordinate, and $\hat V$ is the two-body
nucleon-nucleon interaction. Coupled-cluster calculations starting
from this Hamiltonian have been shown to generate solutions
that are separable into a gaussian center-of-mass wave function and
an intrinsic wave function, see  for example Refs.~\cite{HPD09,Hagen2010a} for further details.

The nucleon-nucleon interaction we use is the chiral N$^3$LO interaction
model of Entem and
Machleidt \cite{Entem2003}.
Although the interaction has a cut-off $\Lambda=500$ MeV, it still
contains high momentum modes, and one typically needs model spaces which
comprise about $20$ major oscillator shells in order to reach convergence for
the ground states of selected oxygen and calcium isotopes, see for
example Refs.~\cite{Hagen2010a,hdhk2010} for a discussion. However,
exploiting the spherical symmetry of the interaction and our
coupled-cluster formalism, we can use model spaces that are large
enough so that there is no need for a subsequent renormalization of the
interaction.

Short-range properties are not modelled explicitly, but are instead
represented by contact terms in the interaction. The contact terms are subject
to the fitting procedure, meaning that the short range dynamics will be sensitive to the
momentum cut-off. Since we are neglecting many-body forces like
three-body forces or more complicated terms, our results will in
general depend, less or more, on the chosen cut-off of the
nucleon-nucleon interaction model. Quantities that are sensitive to
the short-range part of the wavefunctions, such as spectroscopic factors,
may depend strongly on the chosen cut-off. This is why even \emph{ab
  initio} calculations of spectroscopic factors must be considered as
model dependent. Our results may be fully converged in terms of a
given Hamiltonian and its parameters at a given level of many-body
physics, however, employing another nucleon-nucleon interaction may
lead to slightly different results since many-body terms beyond those
represented by a two-body interaction can be very important. This is
discussed in detail in Ref.~\cite{hdhk2010}.

We use a Gamow-Hartree-Fock (GHF) solution for the reference state, as detailed in for example
Ref.~\cite{Hagen2010}. These Hartree-Fock solutions were built from the standard
harmonic oscillator (HO) basis combined with Woods-Saxon (WS) single
particle states for selected partial waves in order to properly reproduce
effects of the continuum. The role of the continuum is expected to be
important close to the drip line, as seen in
Refs.~\cite{Hagen2010,koshiroh2009,michel2009,Volya2006}.
For this purpose we use a Berggren representation \cite{Berggren1968}
for the neutron $s_{1/2}$, $d_{3/2}$, and $d_{5/2}$ partial
waves. This representation generalizes the standard
completeness relation to the complex energy plane. In the Berggren basis, bound,
resonant, and non-resonant continuum states are treated on an equal
footing. The Berggren ensemble has been successfully used within the
Gamow shell model, see for example Ref.~\cite{michel2009} for a
recent review, and in \emph{ab initio} coupled-cluster calculations
of energies and lifetimes in Refs.~\cite{Hagen2007,Hagen2010}.
The Berggren basis is constructed by diagonalizing a one-body Hamiltonian with a
spherical Woods-Saxon potential in a spherical plane-wave basis defined on a
discretized contour in the complex momentum plane. We employ a
total of 30 Gauss-Legendre mesh points along the contour for each of
the $s_{1/2}$, $d_{3/2}$, and $d_{5/2}$ neutron partial waves.
With $30$ discretized single-particle states our results for
the above single-particle states become independent of the choice of contour.
For the choice of interaction that we use, the $1/2^+$ and $5/2^+$
states are fairly well bound with respect to $^{22}$O
\cite{Hagen2010a,Otsuka2010a}, therefore it is sufficient to use a contour along the real
energy axis. For all other partial waves, the basis functions are those of the
spherical harmonic oscillator.

\begin{figure}[thbp]
	\begin{center}
		\includegraphics[width=0.45\textwidth, clip]{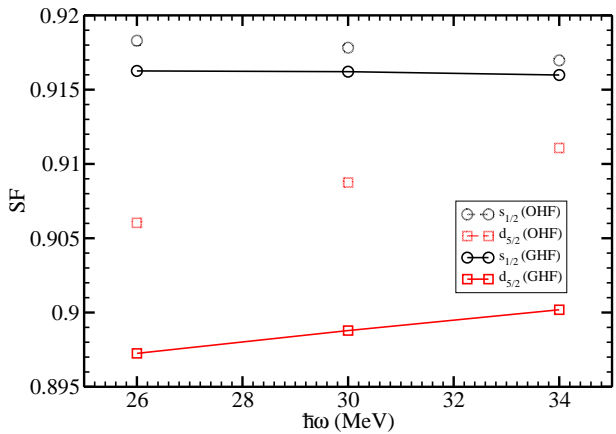}
	\end{center}
	\caption{\label{figSFvshw1}
	(Color online)
	Normalized spectroscopic factors plotted against $\hbar\omega$ for $J^\pi =
	1/2^+$ and $5/2^+$ one-neutron removal from $^{24}$O.  The continuum states
	included in the Berggren-basis calculation (GHF), leads only to a small
	reduction compared with the harmonic oscillator values (OHF).  The dependence
	on the oscillator spacing $\hbar\omega$ is very weak in both calculations.
	The inclusion of the continuum structure also gives a small, but visible,
	improvement in the $\hbar\omega$ dependence.
	}
\end{figure}

\emph{Results} \label{secResults}
Figure \ref{figSFvshw1}  shows the spectroscopic factors for
removing a neutron in the $s_{1/2}$ and $d_{5/2}$ partial waves of
$^{24}$O as function of the harmonic oscillator frequency
$\hbar\omega$. The ground state $1/2^+$ and excited $5/2^+$ state in
$^{23}$O were calculated within the PR-EOM-CCSD approximation starting
from the GHF basis with 30 mesh points for each of the
$s_{1/2}, d_{3/2}$ and $d_{5/2}$ neutron partial waves and 17 major
oscillator shells for the protons and remaining neutron partial
waves. The spectroscopic factors are well converged with respect
to the model space size. To investigate the role of continuum on the
spectroscopic factors, we compare with a calculation done in a
hartree-fock basis built from a harmonic oscillator basis of 17 major
shells (OHF).

\begin{figure}[hbtp]
	\begin{center}
		\includegraphics[width=0.45\textwidth, clip]{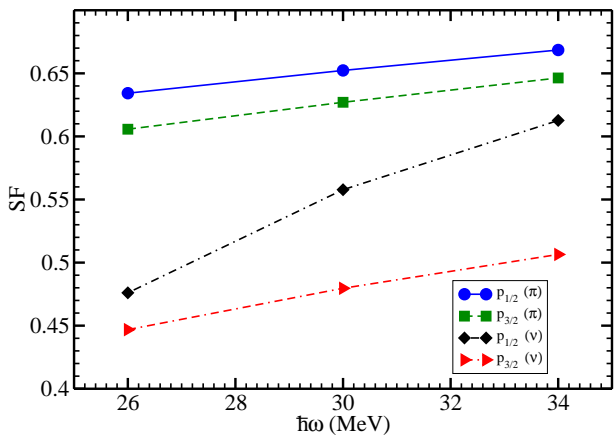}
	\end{center}
	\caption{\label{figSFvshw2}
	(Color online)
	Normalized spectroscopic factors plotted against $\hbar\omega$
        for the negative parity states $J^\pi =
	1/2^-$ and $3/2^-$ in one-proton ($\pi$) and one-neutron ($\nu$)
        removal from $^{24}$O.}
\end{figure}

We find that the
effect of the continuum is small.  This is expected, since our
calculations of the $1/2^+$ and $5/2^+$ single particle states in
$^{24}$O result in  well bound states with respect to the neutron emission
threshold, see for exmaple Ref.~\cite{Hagen2010a} for more details. We do however see
a small reduction of the spectroscopic factors when the  continuum is included. The
reduction of spectroscopic factors will be enhanced for states close to a
reaction channel threshold as discussed in Ref.~\cite{Michel2007a}.
Although the effect of the continuum on the
spectroscopic factors is marginal in the cases we consider, the effect
is crucial in order to obtain the correct asymptotic behaviour of the
overlap functions for one-neutron removal. It is the asymptotic
normalization coefficient, which is calculated from the tail of the radial
overlap function that enters the exact reaction amplitudes.  It is
arguably the relevant quantity to calculate for reactions \cite{Akram10}.
The results in Fig.~\ref{figSFvshw1} show also that the spectroscopic factors for the single neutron
states close to the Fermi surface  are close to one, indicating that a configuration consisting of a single-hole
removal from $^{24}$O captures much of the structure of the wave function for these states. The results for the
neutron $s_{1/2}$ and $d_{5/2}$ hole states lend thus support to the accumulated evidence that
$^{24}$O can be interpreted as a closed-shell nucleus.  For the proton states close to the Fermi surface, this finding
is only partly corroborated by the results shown in Fig.~\ref{figSFvshw2}.
In that figure we show our results for the
spectroscopic factors for removing either a neutron or a proton in
the $p_{3/2}$ and $p_{1/2}$ partial waves of $^{24}$O.
From Fig.~\ref{figSFvshw2} we see that the $3/2^-$ and
$1/2^-$ states in neither $^{23}$O nor $^{23}$N can be clearly interpreted as
simple one-hole states in $^{24}$O. However, the stronger dependence on the
harmonic oscillator frequency $\hbar\omega$ indicates that we
are missing many-body correlations beyond the $2h-1p$ level in
our PR-EOM-CCSD computations of the  $1/2^-$ and $3/2^-$ states.

Our results for spectroscopic factors and energies for the
$J^{\pi} = 1/2^+, 5/2^+, 3/2^-, 1/2^-$ states in $^{23}$O and the
$J^{\pi} = 1/2^-, 3/2^-$ states in $^{23}$N
are summarized in Table~\ref{tableESF}.  We present the values that were
calculated with a harmonic oscillator frequency $\hbar\omega = 30$ MeV, and
the energies are given relative to the ground state of $^{23}$O.
The spectroscopic factors agree with both the extracted value for the
$1/2^+$ state, and the theoretical results obtained with fitted shell model
interactions.  The USDB interaction gives $1.810$ for the neutron $1/2^+$
state and $5.665$ for the neutron $5/2^+$ state.  Corresponding numbers for
SDFP-M  are $1.769$ and $5.593$, respectively \cite{Kanungo2009}.

As can be seen from Table~\ref{tableESF}, our spacing between the two
one-hole states is only $0.35$MeV. This contradicts the shell model
calculations with fitted interactions reported in Ref.~\cite{Kanungo2009}.
The energy spacing is $2.586$MeV and $2.593$MeV for the SDFP-M interaction
and the USDB interaction, respectively.  Assuming a large spectroscopic
factor for the $5/2^+$ state, a bound $5/2^+$-wave state at an energy
consistent with the shell model predictions, should have been seen in the
experiment of Kanungo {\em et al} \cite{Kanungo2009}. The lack of such an
observation was interpreted as a confirmation that this state is unbound.
We speculate that missing
three-nucleon forces could play an important role regarding the shell
gap between the $1/2^+$ and $5/2^+$ single particle states in
$^{24}$O. However, recent  calculations by Otsuka {\em et al} \cite{Otsuka2010a} where
three-body interactions  were included as density-dependent corrected two-body
interactions in a shell-model calculation,
give results for $^{24}$O which are very close to our spacing of
$0.35$ MeV.  Similarly, a many-body perturbation theory calculation using the same Hamiltonian
as here, results in a spacing of $0.39$ MeV. Further investigations are thus needed in order to
clarify the  discrepancies between the results reported in for example
Ref.~\cite{Kanungo2009}, the present results and those of  Otsuka {\em et al} \cite{Otsuka2010a}.
\begin{table}[hbtp]
	\begin{ruledtabular}
	\begin{tabular}{rc|cccc}
		&       & PR-EOM-CCSD  &     &   Experiment & \\
		&       & E     & SF  & E &  SF \\
		\hline
		$^{23}$O& $1/2^+$ & $0.0$  & $1.832$ & $0.0$ & $1.71 \pm 0.19$ \\
		        & $5/2^+$ & $0.35$ & $5.393$ &       &  \\
		        & $3/2^-$ & $12.4$ & $1.919$ &       &   \\
		        & $1/2^-$ & $13.4$ & $1.116$ &       &   \\
		\hline
		$^{23}$N& $3/2^-$ & $20.7$ & $2.609$ &       &   \\
		        & $1/2^-$ & $21.8$ & $1.254$ &  $22.33$    &   \\
	\end{tabular}
\end{ruledtabular}
	\caption{Energies for states in $^{23}$O and $^{23}$N and corresponding spectroscopic
	factors (SF) for the removal of a particle from $^{24}$O.  The reported
	coupled-cluster results are at the level of singles and doubles (CCSD),
	and are calculated with $\hbar\omega=30$MeV.
	Experimental values for $^{23}$O are taken from Ref.~\cite{Kanungo2009} and
	data for $^{23}$N is taken from Ref.~\cite{Audi2003}.
	\label{tableESF}
	}
\end{table}
Compared with the experimental results in Table \ref{tableESF}, we note that the
experimental error bars are typically orders of magnitude larger than the
dependence on technical parameters displayed in Figs.~\ref{figSFvshw1} and \ref{figSFvshw2}.
Finally, we note that the $1/2^-$ state in $^{23}$N is in very good agreement with the theoretical mass evaluations of Ref.~\cite{Audi2003}.

However, there is still a considerable model dependence inherent
in the calculation of spectroscopic factors. First of all, our
calculations have been performed at the level of the singles and
doubles approach, meaning that all correlations up to the level of
two-particle-two-hole excitations are included to infinite order.
Some selected higher $n$-particle-$n$-hole correlations are also
included. The inclusion of triples correlations, that is the admixture of
three-particle-three-hole correlations on the
ground state of the  nucleus with $A$ nucleons and $3h-2p$ excitations in the PR-EOM-CCM
calculations of the $A-1$ nucleus,
on spectroscopic factors remains to be investigated. The largest
uncertainty in our calculations is however most likely the effect of
three-body interactions arising in chiral perturbation theory. A
recent analysis by Otsuka {\em et al} \cite{Otsuka2010a} demonstrates
in shell-model calculations constrained by the degrees of freedom
of the $sd$-shell, that three-body interactions are important in order
to obtain the experimental trend in binding energies for the oxygen
isotopes. Three-body interactions, included as density dependent
corrections to the two-body interactions for the $sd$-shell, result
in $^{25}$O as unbound with respect to $^{24}$O. This is however still
an unsettled topic. A similar effect can be obtained at the level of
two-body effective interactions by including higher-lying
single-particle excitations in many-body perturbation theory.
The role of such three-body interactions in our calculations of spectroscopic factors and
single-particle energies is a topic for future investigations.

\emph{Conclusion and Outlook}
We have computed single-particle energies and spectroscopic factors for hole
states in $^{24}$O using coupled-cluster theory at the level of singles and doubles
correlations. The role of continuum
states has also been included in our investigations. For the hole states the major influence
of the continuum states is to give final single-particle energies and spectroscopic factors
which are almost independent of the chosen oscillator energy.
The spectroscopic factors for protons and neutrons obtained with \emph{ab initio} coupled-cluster
calculations support the emerging consensus that $^{24}$O is a doubly magic
nucleus.
In future work we plan to investigate the application of \emph{ab initio} radial
overlap functions to neutron knock-out reactions on drip line nuclei, thereby removing
a level of model dependence from reaction studies.

\begin{acknowledgments}
Discussions with Gustav R. Jansen are acknowledged.
This work was supported by the Office of Nuclear Physics, U.S. Department of Energy
(Oak Ridge National Laboratory); the University of
Washington under Contract No. DE-FC02-07ER41457.
This research used computational
resources of the National Center for Computational Sciences and the Notur project in Norway.
\O J and MHJ acknowledge support from the Research Council of Norway.
\end{acknowledgments}

\end{document}